\documentclass[twocolumn,preprintnumbers,amsmath,amssymb,floatfix,superscriptaddress,nofootinbib]{revtex4}

\usepackage{amsmath,hyperref,amssymb}
\usepackage{graphicx}
\usepackage{dcolumn}
\usepackage{bm}
\usepackage{verbatim}
\usepackage{tabularx}
\usepackage{slashed}
\usepackage{cancel}
\usepackage[boxsize=2em,aligntableaux=center]{ytableau}
\usepackage{float}
\usepackage{enumerate}
\usepackage{bm}
\usepackage{setspace}
\usepackage{multirow}
\usepackage{hyperref}

\usepackage{enumitem}
\setlist{nolistsep}

\usepackage{float}
\usepackage{placeins}
\usepackage{afterpage}

\def\be{\begin{equation}}
\def\ee{\end{equation}}
\def\bea{\begin{eqnarray}}
\def\eea{\end{eqnarray}}
\newcommand{\eref}[1]{(\ref{#1})}

\newcommand{\vev}[1]{ \left\langle {#1} \right\rangle }

\newcommand{\nn}{\nonumber}

\definecolor{colorTC}{rgb}{.2,.7,.2}
\definecolor{colorDP}{rgb}{0,0,1}
\definecolor{colorRTD}{rgb}{1,0,0}

\begin{document}

\vspace*{-30mm}

\title{$\bm{N}$naturalness}
\author{$\bm{N}$ima~Arkani-Hamed}
\affiliation{\,School of Natural Sciences, Institute for Advanced Study, Princeton, New Jersey 08540, USA\vspace{2pt}}
\author{Timothy~Cohe$\bm{N}$}
\affiliation{\,Institute of Theoretical Science, University of Oregon, Eugene, OR 97403, USA \vspace{2pt}}
\author{Raffaele~Tito~D'Ag$\bm{N}$olo\vspace{3pt}}
\affiliation{\,School of Natural Sciences, Institute for Advanced Study, Princeton, New Jersey 08540, USA\vspace{2pt}}
\author{A$\bm{N}$son~Hook}
\affiliation{\,Stanford Institute for Theoretical Physics, Stanford University, Stanford, CA 94305, USA \vspace{2pt}}
\author{Hyu$\bm{N}$g~Do~Kim}
\affiliation{\,Department of Physics and Astronomy and Center for Theoretical Physics, Seoul National University, Seoul 151-747, Korea\vspace{2pt}}
\author{David~Pi$\bm{N}$ner\,\vspace{3pt}}
\affiliation{\,Princeton Center for Theoretical Science, Princeton University, Princeton, NJ 08544, USA\vspace{5pt}}

\begin{abstract}
\begin{centering}
{\bf Abstract}\\[4pt]
We present a new solution to the electroweak hierarchy problem.
We introduce $N$ copies of the Standard Model with varying values of the Higgs mass parameter. This generically yields a sector whose weak scale is parametrically removed from the cutoff by a factor of $1/\sqrt{N}$.  Ensuring that reheating deposits a majority of the total energy density into this lightest sector requires a modification of the standard cosmological history, providing a powerful probe of the mechanism.  Current and near-future experiments will explore much of the natural parameter space.  Furthermore, supersymmetric completions which preserve grand unification predict superpartners with mass below $m_W \times M_{\text{pl}} / M_{\text{GUT}} \sim 10$~TeV.
\end{centering}
\end{abstract}

\vspace*{1cm}

\maketitle

\begin{spacing}{1.1}

\section{Mechanism}

This letter describes a new mechanism, dubbed ``$N$naturalness,'' which solves the hierarchy problem.  It predicts no new particles at the LHC, but does yield a variety of experimental signatures for the next generation of CMB and large scale structure experiments~\cite{Dodelson:2013pln, Allison:2015qca}.  Well-motivated supersymmetric incarnations of this model predict superpartners beneath the scale $m_W \times M_{\text{pl}}/M_{\text{GUT}} \sim 10$~TeV, accessible to a future 100 TeV collider~\cite{Arkani-Hamed:2015vfh, Golling:2016gvc}.

The first step is to introduce $N$ sectors which are mutually non-interacting.  The detailed particle content of these sectors is unimportant, with the exception that the Standard Model (SM) should not be atypical; many sectors should contain scalars, chiral fermions, unbroken gauge groups, etc.  For simplicity, we imagine that they are exact copies of the SM, with the same gauge and Yukawa structure.  

\begin{figure}[!t]
\centering
\includegraphics[width=0.48\textwidth]{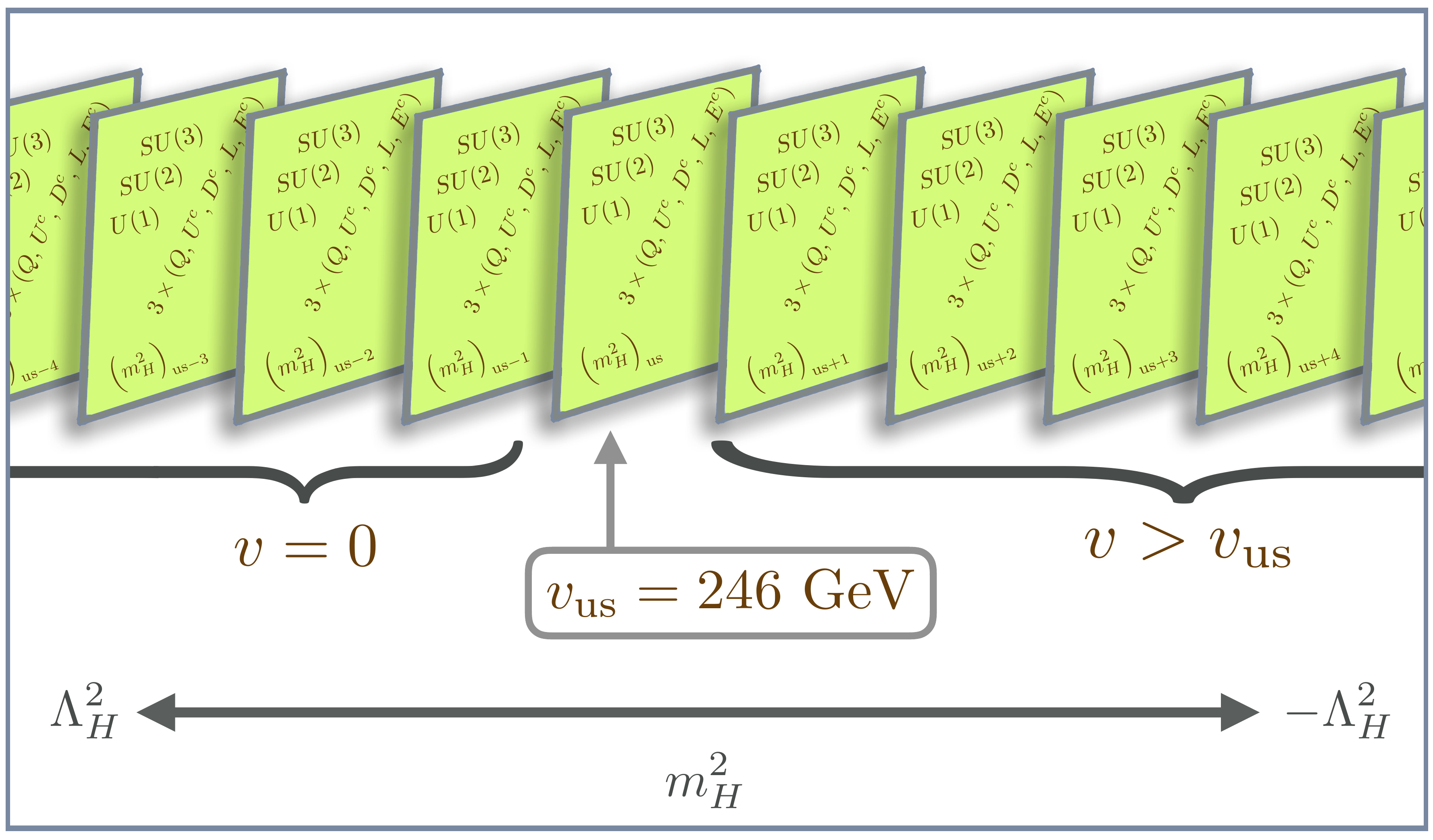}
\caption{A sketch of the $N$naturalness setup.  The sectors have been ordered so that they range from $m_H^2 \sim \Lambda_H^2$ to  $-\Lambda_H^2$.  The sector with the smallest vacuum expectation value contains our copy of the SM.}
\label{fig:Sketch}
\end{figure}

It is crucial that the Higgs mass parameters are allowed to take values distributed between $-\Lambda_H^2$ and $\Lambda_H^2$, where $\Lambda_H$ is the (common) scale that cuts off the quadratic divergences.  Then for a wide range of distributions, the generic expectation is that some sectors are accidentally tuned at the $1/N$ level, $\left|m_H^2\right|_{\text{min}} \sim \Lambda_H^2 / N$.  We identify the sector with the smallest non-zero Higgs vacuum expectation value (vev), $\vev{H} = v$, as ``our" SM.  This picture is illustrated schematically in Fig.~\ref{fig:Sketch}.  

In order for small values of $m_H^2$ to be populated, the distribution of the mass parameters must pass through zero.  For concreteness, we take a simple uniform distribution of mass squared parameters, indexed by an integer label $i$ such that 
\be
\left(m_H^2\right)_i = -\frac{\Lambda_H^2}{N} \big(2\, i+ r \big),\quad\quad  -\frac{N}{2} \leq i \leq \frac{N}{2},
\label{eq:mHsqiScaling}
\ee
where $i=0 = \text{``us"}$ is the lightest sector with a non-zero vev: $\left(m_H^2\right)_\text{us} = -r\times\Lambda_H^2/N  \simeq -(88 \text{ GeV})^2$ is the Higgs mass parameter inferred from observations.  The parameter $r$ can be seen as a proxy for fine-tuning,\footnote{There are a variety of other ways one might choose to implement a measure of fine-tuning in this model.  For example, one could assume the distribution of Higgs mass squared parameters is random with some (arbitrary) prior, and then ask statistical questions regarding how often the resulting theory is compatible with observations.} since it provides a way to explore how well the naive relation between the cutoff and the mass scale of our sector works in a detailed analysis.  Specifically, $r = 1$ corresponds to uniform spacing, while $r<1$ models to an accidentally larger splitting between our sector and the next one.  A simple physical picture for this setup is that the new sectors are localized to branes which are displaced from one another in an extra dimension.  In this scenario, the lack of direct coupling is clear, and the variation of the mass parameters can be explained geometrically: the $m_H^2$ parameters may be controlled by the profile of a quasi-localized field shining into the bulk. 

As a consequence of the existence of a large number of degrees of freedom, the hierarchy between $\Lambda_H$ and the scale $\Lambda_G$ where gravity becomes strongly coupled is reduced.  The renormalization of the Newton constant implies $\Lambda_G^2 \sim M_{\text{pl}}^2 / N$.  If perturbative gauge coupling unification is to be preserved $\Lambda_G \gtrsim M_{\text{GUT}}$, implying that $N \lesssim 10^4$.  This gives a cutoff no greater than $\Lambda_H \sim 10$~TeV, thus predicting a little hierarchy that mirrors the GUT-Planck splitting in the UV.  At the scale $\Lambda_H$, new dynamics (\emph{e.g.}, SUSY) must appear to keep the Higgs from experiencing sensitivity to even higher scales.   Alternatively, the full hierarchy problem can be solved with $N \sim 10^{16}$, so that $\Lambda_H \sim \Lambda_G \sim 10^{10}$~GeV.  Note that this number of copies, while sufficient, is unnecessary for a complete solution.  There may be two classes of new degrees of freedom: the $N$ copies that participate directly in the $N$naturalness picture, and another completely sterile set of degrees of freedom that still impact the renormalization of $\Lambda_G$.

So far we have described a theory with a $S_N$ permutation symmetry, broken softly by the $m_H^2$ parameters, such that each of the sectors is SM-like. 
Sectors for which $m_H^2 < 0$ are similar to our own, with the exception that particle masses scale with the Higgs vev, $v_i \sim v\,\sqrt{i}$.  In addition, once $i \gtrsim 10^{8}$ the quarks are all heavier than their respective QCD scales.  Those sectors do not exhibit chiral symmetry breaking, nor do they contain baryons. Sectors with $m_H^2>0$ are dramatically different from ours.  In these sectors, electroweak symmetry is broken at low scales due to the QCD condensate $\Lambda_\text{QCD}$.  Fermion masses are generated by the four-fermion interactions that are induced by integrating out the complete $SU(2)$ Higgs multiplet.  Thus, $m_f \sim y_f\,y_t\,\Lambda_{\text{QCD}}^3 / \left(m_{H}^2\right)_i \lesssim 100$~eV, where $y_t$ is the top Yukawa coupling.  All fermionic and gauge degrees of freedom are extremely light relative to the ones in our sector.

With so many additional degrees of freedom, the naive cosmological history is dramatically excluded.  In particular, if all sectors have comparable temperatures in the early Universe, then one expects $\Delta N_{\text{eff}} \sim N$ (see Eq.~(\ref{eq:dNeff})).  Thus, the hierarchy problem gets transmuted into the question of how to predominantly reheat only those sectors with a tuned Higgs mass.

To accomplish this, we need to introduce a last ingredient into the story, the ``reheaton" field, so named because it is responsible for reheating the Universe via its decays.  We call this field $S^c$ for models where the reheaton is a fermion, and $\phi$ if the reheaton is a scalar.  The cosmological history of the model begins in a post-inflationary phase where the energy density of the Universe is dominated by the reheaton. As stated multiple times we can not be unique, therefore we assume that the reheaton couples universally to all sectors.  Note that the scalars must be near their true minimum when reheating occurs.  This can be accomplished by having either low scale inflation, or else a coupling of the Higgses to the Ricci scalar.

In the next section, we present a set of models in which the reheaton dynamically selects and populates only the lightest sectors, despite preserving the aforementioned softly broken $S_N$ symmetry.  Sec.~\ref{sec:Signals} then provides constraints on these models, and Sec.~\ref{sec:Discussion} contains our conclusions and highlights potential signals.  

\section{Models}
\label{Models}
We have argued that the hierarchy problem can be solved by invoking a large number of copies of the SM, along with some dynamical mechanism which dominantly populates the lightest sector with a non-zero Higgs vev.  This section details some simple explicit models that realize a viable cosmological history.  

As anticipated in the previous section, we imagine that at a post-inflationary stage the energy density of the Universe is dominated by a reheaton that couples universally to all the new sectors. Its decays populate the SM and its copies. The goal is to deposit as much energy as possible into the sector with the smallest Higgs vev.   This may be accomplished by arranging the decays of the reheaton such that the branching fraction into the $i^\text{th}$ sector scales as $\text{BR}_i \sim \left(m_H\right)_i^{-\alpha}$ for some positive exponent $\alpha$.  To this end, we construct models that share three features:
\vspace{2pt}
\begin{enumerate}[label=(\roman*)]
\item The reheaton is a gauge singlet; 
\vspace{3pt}
\item It is parametrically lighter than the naturalness cutoff, $m_\text{reheaton}\lesssim \Lambda_H/\sqrt{N}$;
\vspace{3pt}
\item Its couplings are the most relevant ones possible that involve the Higgs boson of each sector.  
\end{enumerate}

While the requirement of a light reheaton field may appear to require an additional coincidence, it can be easily accommodated in an extra-dimensional picture.  In order to couple to all the sectors, the reheaton must be a bulk field.  Then, before canonical normalization, its kinetic term carries a factor of $N$.  If the reheaton enjoys a shift symmetry that is respected in the bulk, it will receive a $\Lambda_H$-sized mass from each brane on which the shift symmetry is violated.  Here we assume that the dynamics above $\Lambda_H$ respect the shift symmetry.  As long as the shift symmetry is only violated on the boundaries, the reheaton mass will be parametrically the same as the weak scale after canonical normalization.  In the case of a fermionic reheaton, this simple picture corresponds to the brane-localization of its Dirac partner.

The two simplest models, which we denote $\ell$ and $\phi$, are 
\be
\mathcal{L}_\ell \supset -\lambda\,S^c\,\sum_i \ell_i\,H_i - m_S\,S\, S^c\,,
\ee  
if the reheaton is a fermion $S^c$, and 
\begin{align}
\mathcal{L}_\phi \supset -a \,\phi\,\sum_i \,|H_i|^2 -\frac{1}{2}\, m^2_\phi\,\phi^2,
\end{align} 
if the reheaton is a scalar $\phi$.  For the theory to be perturbative, we need the coupling $\lambda$ to obey a `t Hooft-like scaling $\lambda \sim 1/\sqrt{N}$. Naively we would expect the same scaling for $a$, but we find that a stronger condition needs to be imposed ($a \sim 1/N$) to insure that the loop induced mass for $\phi$ is not much larger than $\Lambda_H/\sqrt{N}$.  Even with this scaling, the loop-induced tadpole for $\phi$ will be too large unless the sign of $a$ is taken to be arbitrary for each sector.  Note that $a$ breaks a $\mathbb{Z}_2$ symmetry on $\phi$, so that this choice is consistent with technical naturalness.  Including the arbitrary sign, the sum over tadpole contributions only grows as $\sqrt{N}$, and so the natural range of $\phi$ is restricted to $\Lambda_H \sqrt{N}$.  The Higgses will then receive a contribution to their $m_H^2$ parameters of order $a \langle \phi \rangle \sim \Lambda_H^2 / \sqrt{N}$.  While these contributions may be large compared to our weak scale, as long as they are smaller than $\mathcal{O}(\Lambda_H^2)$, they can be safely absorbed into the quadratically-divergent contributions to $m_H^2$.  Of course, these are upper bounds on the couplings; as we will discuss later in the section, they can be consistently taken smaller, so long as the reheat temperature is sufficiently high.

Before moving on to discuss the details of reheating, we remark on the existence of cross-quartics of the form $\kappa\,|H_i|^2 \,|H_j|^2$.  Even if these are absent in the UV theory, they will be induced radiatively.  After electroweak symmetry breaking in the various sectors, these can potentially affect the spectrum, and so it is critical to the $N$naturalness mechanism that they be sufficiently suppressed.  Given an arbitrary, $S_N$ symmetric cross-quartic, $\kappa$, the $m_H^2$ parameters will shift by approximately $ -\kappa\, \Lambda_H^2 \,N / 8 + \mathcal{O} (\kappa^2 N)$, while the mixing effects are subdominant.  Thus, the general picture of hierarchical weak scales remains intact so long as $\kappa \lesssim 1 / N$.

At a minimum, cross-quartics of this form will be induced gravitationally, regardless of the reheaton dynamics.  These quartically-divergent gravitational couplings arise at three loops, giving $(16 \,\pi^2)^3 \kappa_g \sim \lambda_h^2 (\Lambda_H / M_{\text{pl}})^4 \sim (\lambda_h / N)^2 (\Lambda_H / \Lambda_G)^4$, where $\lambda_H$ is the SM-like Higgs self quartic.  Here we have taken the scale that cuts off these divergences to be $\Lambda_H$, as would be appropriate for a supersymmetric UV completion (for which these quartics are absent).  In either case, these gravitational couplings are parametrically safe, since they scale as $(1 / N)^2$.  

In addition, potentially dangerous cross-quartics can be generated by reheaton exchange.  In the $\ell$ model, the cross-quartic is generated at one loop: $\kappa_{\ell} \sim \lambda^4 / 16 \,\pi^2 \lesssim 1 / N^2$, after enforcing the large-$N$ scaling of $\lambda$.  In the $\phi$ model, these quartics are generated at tree-level, $\kappa_{\phi} \sim a^2 / m_{\phi}^2$.   Naively this appears borderline problematic, since $\kappa_{\phi}$ scales as $1 / N$.  However, the arbitrary sign of $a$, which was necessary to mitigate the tadpole of $\phi$, will once again soften the sum over sectors, so that $\sum a_i \,v_i^2 \sim a\, \Lambda_H^2\, \sqrt{N}$.  Combined with the large-$N$ scaling of $a$, these quartics are rendered safely negligible.

\subsection{Reheating}
If the reheaton is sufficiently light, then we may analyze the leading reheaton decay operators using an effective Lagrangian computed by integrating out $H_i$.  This immediately makes it clear why we we want the reheaton to be coupled with the most relevant coupling possible, since these will suffer the fastest suppression as $|m_H|\rightarrow \infty$.  Integrating out the Higgs and gauge bosons in the $\ell$ model, the leading decays of $S^c$ are given by, \emph{e.g.}
\be
\begin{array}{l}
\mathcal{L}^{\vev{H}\neq 0}_\ell \supset \mathcal{C}^\ell_1\, \lambda\,\frac{v}{m_Z^2 \,m_S}\nu^\dag \bar{\sigma}^\mu S^c \, f^\dagger \bar{\sigma}_\mu f \,;\\[10pt]
\mathcal{L}^{\vev{H}= 0}_\ell \supset \mathcal{C}^\ell_2\,\lambda\,\frac{y_t}{m_H^2}S\, \ell\, Q_3^\dag \,u^{c\dag}_3\,,
\end{array}
\label{Eq: fermion Dim 1}
\ee
where $m_{Z}$ is the relevant $Z^0$-boson mass and the $\mathcal{C}^\ell_i$ are numerical coefficients.  We have omitted decays through $W$ and Higgs bosons for sectors with $\vev{H} \neq 0$ as they scale in the same way. We include them in all numerical computations.

From this low energy Lagrangian we can easily infer that a light reheaton dominantly populates the lightest negative Higgs mass sector. Denoting with $m_{h_i}$ the physical Higgs mass in sectors with $\langle H\rangle \neq 0$, the reheaton decay widths scale as $\Gamma_{m_H^2<0} \sim 1/m_{h_i}^2$ and $\Gamma_{m_H^2>0} \sim 1/m_{H_i}^4$ in sectors with and without electroweak symmetry breaking, respectively.  Thus the reheaton preferentially decays into sectors with light Higgs bosons and non-zero vevs.   If, instead, the reheaton were heavy enough to decay directly to on-shell Higgs or gauge bosons, the branching fractions would be democratic into those sectors, and the energy density in our sector would not come to dominate the energy budget of the Universe. 

\begin{figure}[t!]
\centering
\includegraphics[width = .45\textwidth]{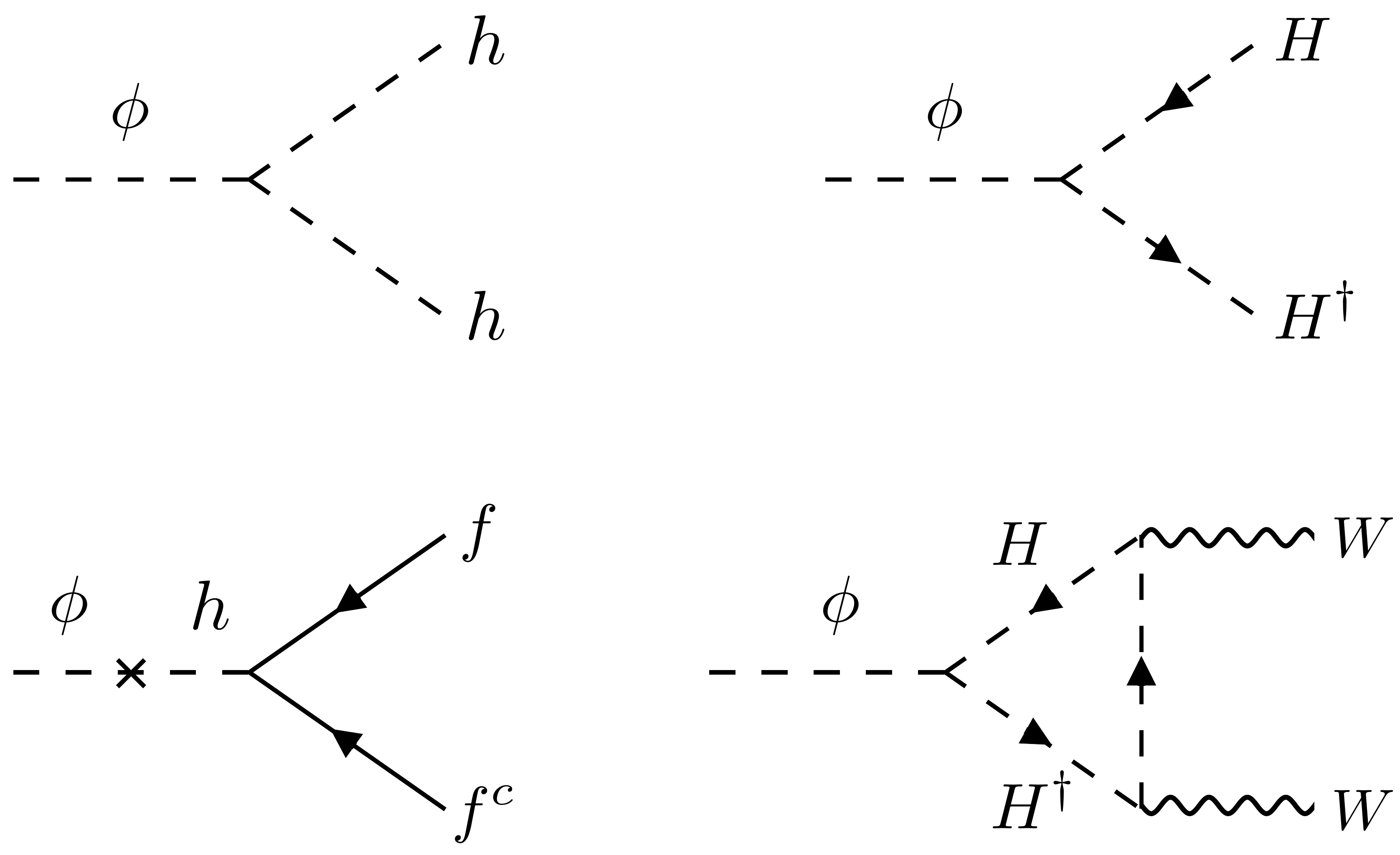} 
\caption{Feynman diagrams for the most important decays in the $\phi$ model.  The left (right) column is for $\vev{H} \neq 0$ $\big(\!\vev{H} = 0\big)$.  The top (bottom) row is for $m_\phi \gg |m_H|$ $\big(m_\phi \ll |m_H|\big)$.   }
\label{fig:PhiDiagrams}
\end{figure}

In the scalar case the decays are different, but the scaling of the decay widths is exactly the same. This can be seen once more by integrating out the Higgs and gauge bosons in all the sectors:
\be
\begin{array}{l}
\mathcal{L}^{\vev{H}\neq 0}_\phi \supset \mathcal{C}_1^\phi\, a\,y_q\frac{v}{m_h^2} \phi\,q\, q^c \,;\\[10pt]
\mathcal{L}^{\vev{H}= 0}_\phi \supset \mathcal{C}_3^\phi\,a\,\frac{g^2}{16\,\pi^2}\frac{1}{m_H^2}\,\phi\, W_{\mu\nu}W^{\mu\nu} \,,
\end{array}
\label{Eq: scalar Dim}
\ee
where again the $\mathcal{C}^\phi_i$ are numerical coefficients, and $W_{\mu\nu}$ is the $SU(2)$ field strength. As in the fermionic case, this Lagrangian leads to decay widths that scale as $\Gamma_{m_H^2<0} \sim 1/m_{h_i}^2$ and $\Gamma_{m_H^2>0} \sim 1/m_{H_i}^4$ in sectors with and without electroweak symmetry breaking, respectively, through the diagrams shown in Fig.~\ref{fig:PhiDiagrams}.  We have not included the one-loop decay $\phi \rightarrow \gamma \,\gamma$ in Eq.~\eref{Eq: scalar Dim} for sectors with $\vev{H}\neq0$.  This operator scales as $1/m_h^2$ and is important for sectors with $N \gtrsim 10^8$; we find that this is never the leading decay once the bounds on $N$ discussed in Sec.~\ref{sec:Signals} are taken into account.

Before moving to a more detailed discussion of signals and constraints it is worth pointing out two important differences between the $\phi$ and $\ell$ models that will lead us to modify the latter. Given the scaling of the widths we can approximately neglect the contributions to cosmological observables from the $\vev{H} = 0$ sectors.  In the simple case that the vevs squared are equally spaced, $v^2_i \sim 2\, i\times v_\text{us}^2$, as in Eq.~(\ref{eq:mHsqiScaling}) with $r=1$, we find that the branching ratio into the other sectors is $\sum 1/i \sim \log N$.

\begin{figure*}[t!]
\centering
\includegraphics[width = .47\textwidth]{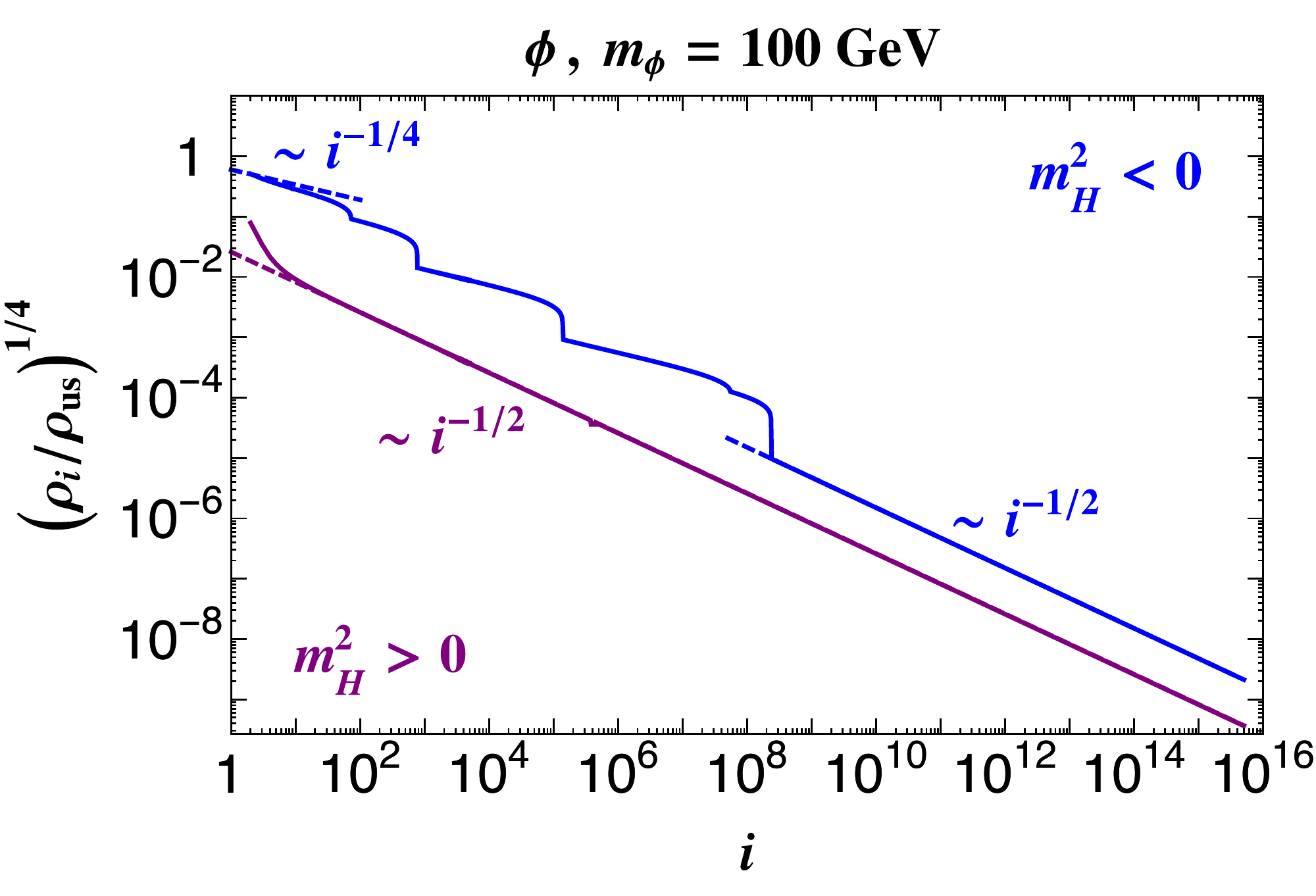} \quad\quad
\includegraphics[width = .47\textwidth]{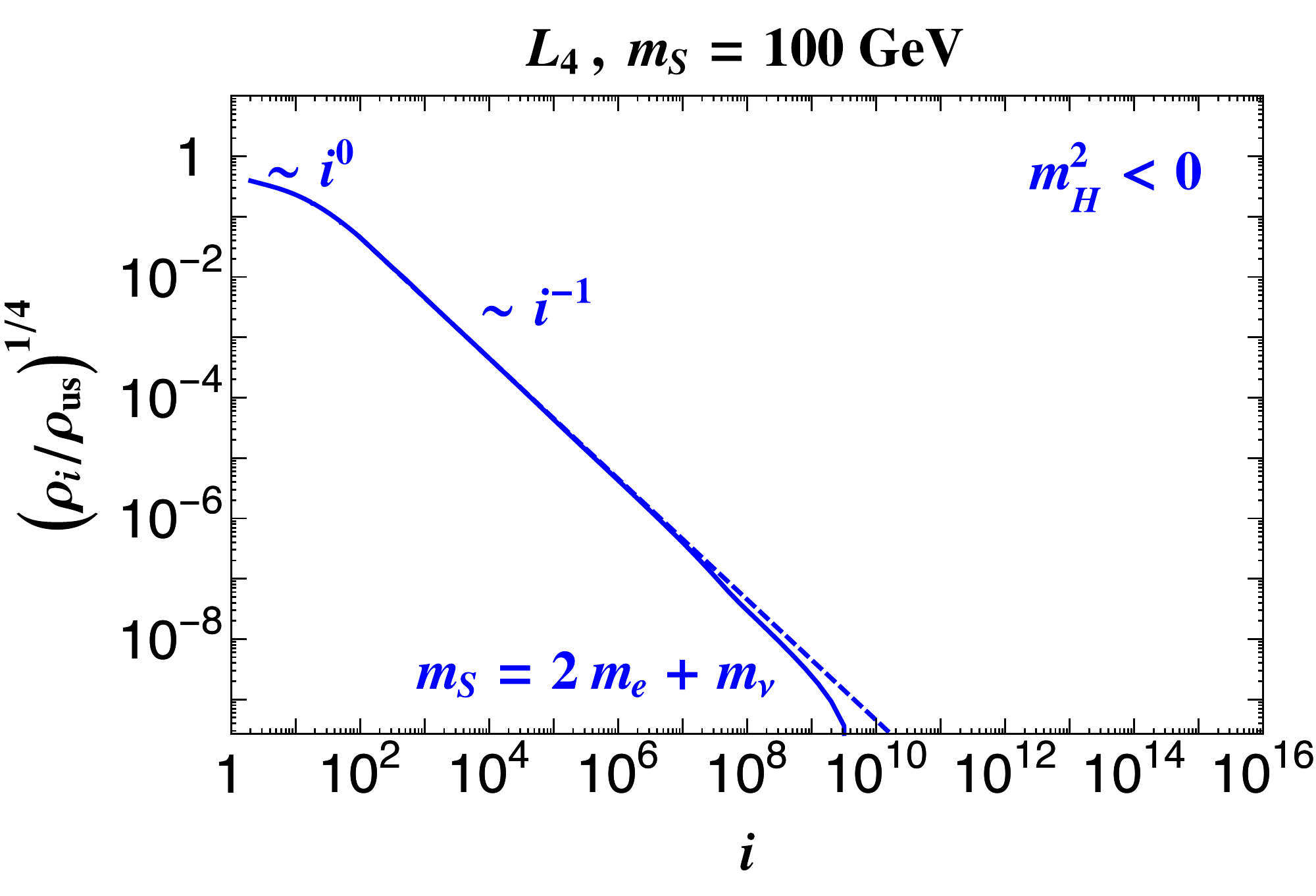}
\caption{Energy density deposited in each sector as a function of sector number, normalized to the energy density in our sector. The left panel is for the $\phi$ model with $a=1$~MeV.  The right panel is for the $L_4$ model with $\lambda\times \mu_E=1$~MeV, $M_L=400$~GeV, $M_{E, N}=500$~GeV, $Y_E=Y_N=0.2$, and $Y^c_E=Y^c_N=-0.5$. The solid lines are the result of a full numerical calculation. The dashed lines show the expected scalings. As discussed in the text, the steps in the $\phi$ model are proportional to Yukawa couplings due to the fact that $\phi$ decays via mixing with the Higgs. When $i \gtrsim 10^9$ in the $L_4$ model, the process $S^c \rightarrow 2\,e+\nu$ cannot proceed on-shell, which results in the deviation from the naive scaling as denoted by $m_S = 2\,m_e + m_\nu$. Both figures were made using the zero temperature branching ratios of the reheaton; thermal corrections are under control so long as $T_\text{RH}$ is smaller than the weak scale in our sector, as discussed at the end of Sec.~\ref{Models}.}
\label{fig:TRH}
\end{figure*}

In the $\phi$ model, this logarithmic sensitivity to $N$ is not realized.  Since the reheaton decays into sectors with non-zero vevs via mixing with the Higgs, the decays become suppressed by smaller and smaller Yukawa couplings as $h_i$ becomes heavy.  After the charm threshold is crossed $m_\phi < 2 \,m_{c_i}$ we can neglect the contribution of the new sectors to cosmological observables (with one exception that we discuss in the next section). This behavior is displayed in the left panel of Fig.~\ref{fig:TRH}, where we show the fraction of energy density deposited in each sector.

The second important difference is that in the $\ell$ model the reheaton couples directly to neutrinos and, in the sectors with electroweak symmetry breaking, it mixes with them.  This leads to two effects.  First, the physical reheaton mass grows with $N$, implying that the structure of the $\ell$ model forces the reheaton to be heavy at large $N$, and can be inconsistent depending on the value of $\lambda$.  Additionally, this mixing can generate a freeze-in abundance~\cite{Hall:2009bx} of neutrinos in the other sectors from the process $\nu_\text{us}\,\nu_\text{us} \rightarrow \nu_\text{us}\,\nu_i$ via an off-shell $Z^0$.  Tension with neutrino overclosure and overproduction of hot dark matter leads to an upper bound on the maximum number of sectors. In practice, it is hard to go beyond $N\simeq 10^3$.

However, there is a simple extension of the $\ell$ model that at once mitigates its UV, \emph{i.e.}, large $N$, sensitivity and solves the problems arising from a direct coupling to neutrinos.  If the reheaton couples to each sector only through a massive portal (whose mass grows with $v_i$), then the branching ratios will scale with a higher power of the Higgs vev after integrating out the portal states.  As an example, consider introducing a 4$^\text{th}$ generation of vector-like leptons $(L_4, L^c_4)$, $(E_4, E^c_4)$, and $(N_4, N^c_4)$ to each sector.  Then relying on softly broken $U(1)$ symmetries, we can couple the reheaton to $L_4$ only via the Lagrangian

\begin{align}
&\mathcal{L}_{L_4} \supset \mathcal{L}_{\rm mix} +\mathcal{L}_{Y} + \mathcal{L}_{M}\, ,  \\
&\mathcal{L}_{\rm mix} = -\lambda\,S^c\, \sum_i \big(L_{4}\,H\big)_i - \mu_E\, \sum_i \big(e^c\, E_4\big)_i\; , \nn \\
&\mathcal{L}_{Y} = -\sum_i\Big[Y_E \big(H^\dag\,L_4\,E^c_4\big)_i + Y_E^c \big(H\,L^c_4\,E_4\big)_i \nn\\
&\qquad\qquad\qquad +Y_N \big(H\,L_4\,N_4^c\big)_i + Y_N^c \big(H^\dag\,L_4^c\,N_4\big)_i \Big]\, , \nn \\
&\mathcal{L}_{M} = -\sum_i\Big[M_E \big(E_4^c\,E_4\big)_i  + M_L \big(L_4^c\, L_4\big)_i \nn \\
&\qquad\qquad\qquad+ M_N\big(N_4^c\,N_4\big)_i\Big]  - m_S\, S \,S^c\,,\nn
\label{eq:L4lagr}
\end{align}
where we have assumed universal masses and couplings across all the sectors for simplicity. We again need $\lambda\sim 1/\sqrt{N}$ for perturbativity. Note that we are assuming that the bilinear $\mu_E\,e^c\,E$ only couples a single flavor of right handed lepton to the new 4$^\text{th}$ generation fields, in order to avoid flavor violation bounds in the charged lepton sector.  The predictions relevant to cosmology (see Fig.~\ref{fig:Neff}) are insensitive to the choice of flavor; we choose couplings involving the $\tau$ for the additional constraints discussed in Sec.~\ref{sec:MixingBetweenSectors} below since this choice yields the strongest bounds.

To explore the differences between the $L_4$ and $\ell$ models let us again consider the limit in which the reheaton is light. If we integrate out the Higgs and gauge bosons along with the new vector-like leptons, the leading operators for the decays of $S^c$ are given by
\be
\begin{array}{l}
\!\!\!\!\mathcal{L}^{\vev{H}\neq 0}_{L_4}\! \supset \mathcal{C}^{L_4}_1 \lambda'\,\frac{g^2}{m_W^2}\, \Big(e^{c\dag}\bar{\sigma}^\mu S^c\Big)\Big(f^\dagger \bar{\sigma}_\mu f^\prime\Big)\,;\\[10pt]
\!\!\!\!\mathcal{L}^{\vev{H}= 0}_{L_4}\! \supset \mathcal{C}^{L_4}_2\lambda\,\frac{y_t \,y_b}{16\,\pi^2}\frac{Y_E \,M_E\, \mu_E}{m_H^4}\,\Big(e^{c\dagger}\bar{\sigma}^{\mu} S^c\Big)\!\Big(u^{c\dagger}_3 \bar{\sigma}_{\mu} d^c_3\Big) \,,
\end{array}
\label{Eq:L4eff}
\ee
where once more the $\mathcal{C}^{L_4}_i$ are numerical coefficients, $M_4$ is used to represent the physical mass of the relevant heavy lepton, and for convenience we have defined $\lambda'_i \equiv \left(\lambda \,v^2_i\,\mu_E/ M_{4i}^4\right) f(Y, M)$. Here $f$ is a function of dimension one that depends on the Yukawa couplings and vector-like masses in Eq.~(\ref{Eq:L4eff}), but not on the Higgs vev.  The $M_{4 i}$ masses receive a contribution from $v_i$ that eventually dominates. When this happens $S^c$ decays become suppressed by large powers of the Higgs vev. From the effective Lagrangian above, it is easy to conclude that the widths scale as $\Gamma_{m_H^2<0}\sim {\rm const}$ for the first few sectors, since $M_{4 i}$ is approximately independent of $v_i$.  When the Yukawa contribution to the masses begins to dominate, such that $M_{4i} \sim v_i$, the scaling becomes $\Gamma_{m_H^2<0}\sim 1/v_i^8$. Contributions to observables from the sectors with positive Higgs mass squared are negligible: the decay is both three-body and loop-suppressed, and the width scales as $1/v_i^8$ in all the sectors. 

\begin{figure}[!t]
\centering
\includegraphics[width = .48\textwidth]{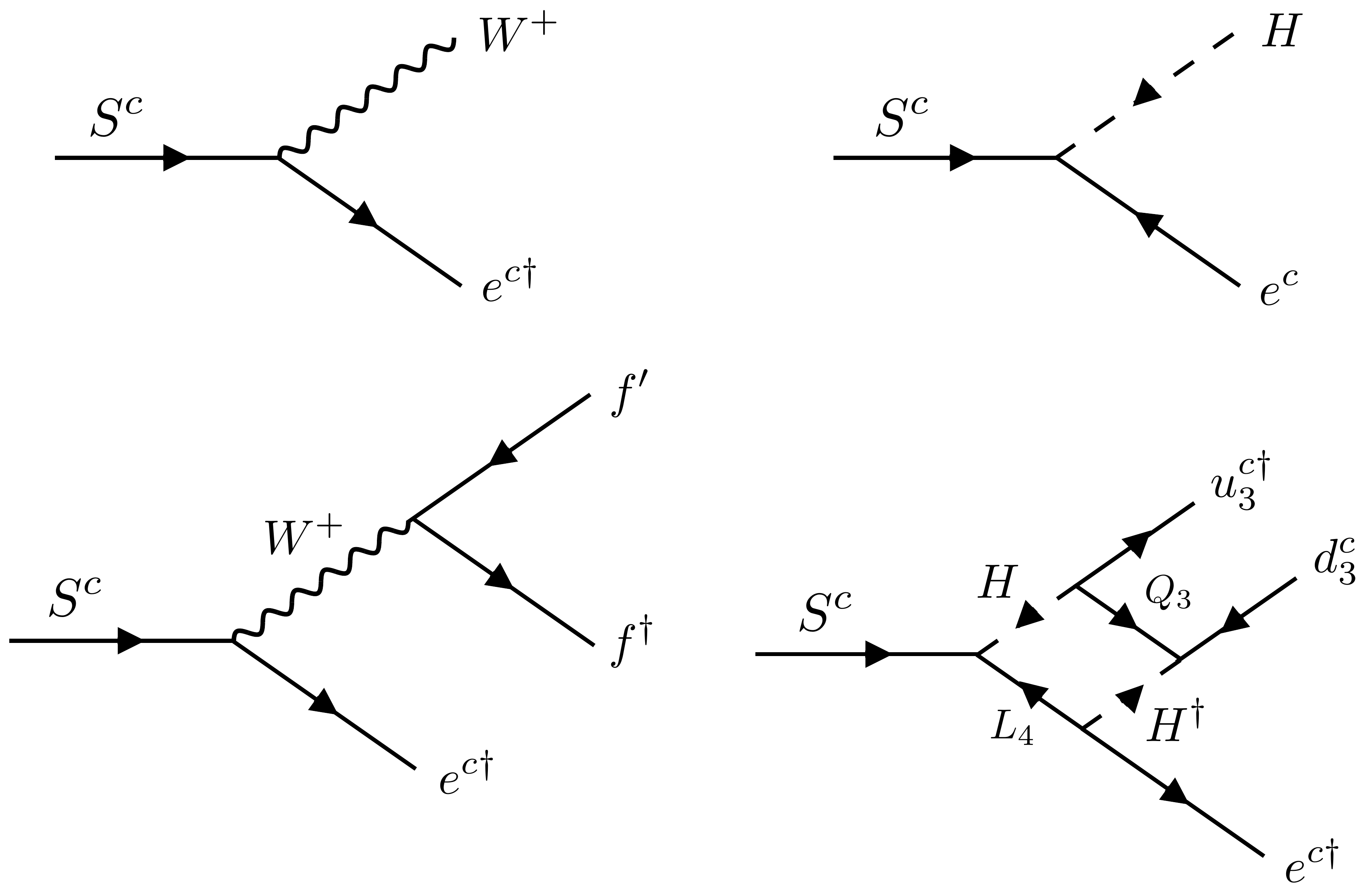} 
\caption{Feynman diagrams for the most important decays in the $L_4$ model.  The left (right) column is for $\vev{H}\! \neq\! 0\, \big(\!\!\vev{H}\! =\! 0\big)$.  The top (bottom) row is for $m_S \gg |m_H|$ $\big(m_S \ll |m_H|\big)$.}
\label{fig:L4Diagrams}
\end{figure}

The diagrams that lead to these decays are shown in Fig.~\ref{fig:L4Diagrams}, and the energy density deposited in each sector is depicted in the right panel of Fig.~\ref{fig:TRH}.  It is obvious that in this model cosmological observables are sensitive only to the few sectors for which the vector-like masses dominate over the Higgs vev, making it insensitive to the UV. This comes at the price of introducing new degrees of freedom near the weak scale.  As we will discuss in the following section, the vector-like masses cannot be arbitrarily decoupled, but they must be large enough to avoid tension with direct searches and the measured properties of our Higgs.

Finally, we end this section by briefly commenting on the presence of an upper bound for the reheating temperature $T_\text{RH}$ such that the mechanism is preserved.  Specifically, $T_\text{RH}$ should be at most of order of the weak scale.  If the temperature were larger, our Higgs mass would be dominated by thermal corrections resulting in a change in the scalings of the branching ratios.  Our Higgs would obtain a large positive thermal mass and no longer be preferentially reheated over the other sectors.  Noting that
\bea
T_\text{RH} \simeq 100 \text{ GeV } \sqrt{\frac{\vev{\Gamma_\text{reheaton}}_T}{10^{-14} \text{ GeV}}},
\label{eq:TRH}
\eea
where $\vev{\Gamma_\text{reheaton}}_T$ denotes a thermal average of the reheaton width that incorporates the effect of time dilation.  Then Eq.~\eref{eq:TRH} places an upper bound on the couplings of the reheaton.  In the $\phi$ model, the $\phi - h$ mixing angle is bounded to be $\theta_{\phi h} \sim \left(a \,v / m_h^2\right)_{\text{us}} \lesssim 10^{-6}~\left(100\text{ GeV} / m_{\phi}\right)^{1/2}$.  In the $L_4$ model, most of the viable region of parameter space predicts on-shell decays to our $W$ boson (see Fig.~\ref{fig:Neff} below).  Therefore, the width of $S^c$ is dominated by this two-body decay and the constraint on $T_\text{RH}$ translates into a rough bound of $\lambda'_\text{us} \lesssim 10^{-7}$  when $m_S \simeq 100\text{ GeV}$.  For the benchmark values used for the figures below, this in turn translates into a bound $\lambda\times \mu_E \lesssim 10^{-2}~\text{GeV}$.  

Finally, we note that at large $N$ there is a more stringent upper bound on the reheating temperature determined by the perturbativity of $\lambda$.  Requiring $\lambda~\lesssim~4\,\pi/\sqrt{N}$ and $m_S~\sim~100$~GeV, we find that it is still possible to reheat to a few GeV even with $N \sim 10^{16}$, where this estimate has been done using the complete numerical implementation of the mixings.

In principle, we must also ensure that other sectors are not overly heated by scattering from our own plasma after reheating.  However, the aforementioned constraints on the reheaton couplings sufficiently suppress this contribution to their energy density.

\subsection{Baryogenesis}
A viable mechanism for baryogenesis is an even more crucial part of our mechanism for solving the hierarchy problem than in typical natural theories for new physics, where it can be treated in a modular way. One challenge is that our reheating temperature should be near or below the electroweak phase transition. Additionally, baryogenesis cannot occur in all of the copies of the SM, or there would be too much matter in the Universe. 

One simple approach, which makes use of features intrinsic to the model, is to imagine that the reheaton $S^c$ carries a lepton number asymmetry.  This asymmetry is distributed to the various sectors through the decays of $S^c$.  Only in the sectors nearest ours is this lepton asymmetry converted into a baryon asymmetry.  The small number abundance of baryons results from the low reheat temperature.  At temperatures just below the electroweak phase transition, the sphaleron rate is exponentially suppressed, and only a small fraction of the lepton asymmetry is converted into a baryon asymmetry.  The baryon asymmetry in sectors with $m_H^2 > 0$ is even further suppressed; since $m_W \lesssim \Lambda_{\text{QCD}}$, the sphalerons remain active at temperatures below the baryon masses.  Any asymmetry in these sectors will eventually be redistributed back into the leptons.  We have now laid out the necessary ingredients of our mechanism and we are ready to explore their phenomenology in more detail. 

\section{Signals and Constraints}
\label{sec:Signals}
The signals and experimental constraints for $N$naturalness come from two sources: mixing between the sectors and energy density deposited in the new sectors by the reheaton decays.   The cosmological observables sensitive to the energy density in each sector can be further divided into two categories.  

First we discuss measurements that can detect new light particles. These signatures are dominated by the sectors closest to us and  can not be avoided by changing the UV scalings of the model. They provide the most characteristic signatures of the theory.  Then we study the impact of stable massive particles from the new sectors. This last set of constraints is dominated by sectors with the largest Higgs masses and can be ignored in the $L_4$ model, where the large $i$ physics is decoupled.  In the last two subsections we discuss the bounds arising from mixing between the sectors, followed by possible collider signatures.

\subsection{Massless degrees of freedom}
\label{subsec:massless}
As discussed previously, our models have a large number of massless or nearly massless degrees of freedom. For example, all additional sectors contain photons and neutrinos. There are several kinds of cosmological observations that are sensitive to new relativistic particles. For instance the measurement of the Hubble parameter during either Big Bang Nucleosynthesis or at the epoch of photon decoupling, and bounds on hot dark matter from the matter power spectrum.   

The sensitivity of the expansion of the Universe to new relativistic degrees of freedom is usually phrased in terms of the number of effective neutrinos
\be
\Delta N_{\rm eff} = \frac{1}{\rho_\nu^\text{us}} {\displaystyle \sum_{i\neq \text{us}} \rho_i}\, .
\label{eq:dNeff}
\ee
Current bounds are $\Delta N_{\rm eff} \lesssim 1$ during BBN~\cite{Cooke:2013cba} and $\Delta N_{\rm eff} \lesssim 0.6$ at photon decoupling~\cite{Ade:2015xua}. In both cases we quote an approximate $95\%$ C.L. constraint. The CMB bound applies to free-streaming radiation~\cite{Bell:2005dr}. However, the photons in some of the new sectors are still in equilibrium with or have just decoupled from electrons at that time and might be more similar to a perfect fluid. Until recently it was impossible to distinguish between the two types of radiation, as they affect the CMB damping tail in the same way~\cite{Hou:2011ec}. The detection of a phase shift in the CMB anisotropies~\cite{Follin:2015hya} has broken this degeneracy, and it is now possible to set a $95\%$ C.L. bound: $N_{\rm fluid}\lesssim 1$ for $\Delta N_{\rm eff}=0$~\cite{Baumann:2015rya}. Here we have defined $N_{\rm fluid}$ in the same way as $\Delta N_{\rm eff}$, normalizing the energy density of non-free-streaming radiation to that of a neutrino in our sector. 

In the following, we do not distinguish between the two types of radiation. We use $\Delta N_{\rm eff}$ to denote the sum of the two components. Given the bounds discussed above and the two dimensional exclusions in~\cite{Baumann:2015rya}, this is sufficient to show that the model has large areas of parameter space consistent with current data. In the future, it would be interesting to explore CMB observations in more detail, as it is a generic prediction of this type of theories to have roughly comparable amounts of free-streaming and non-free-streaming extra radiation.

Having set $N_{\text{fluid}}$ to zero, it is straightforward to estimate the contribution to $\Delta N_{\rm eff}$ from our new sectors, since the ratio of energy densities $\rho_i/\rho_\text{us}$ is determined by the decay widths of the reheaton: $\rho_i/\rho_\text{us}\simeq\Gamma_i/\Gamma_\text{us}$. 
\begin{figure*}[!t]
\begin{center}
\includegraphics[width=0.47\textwidth]{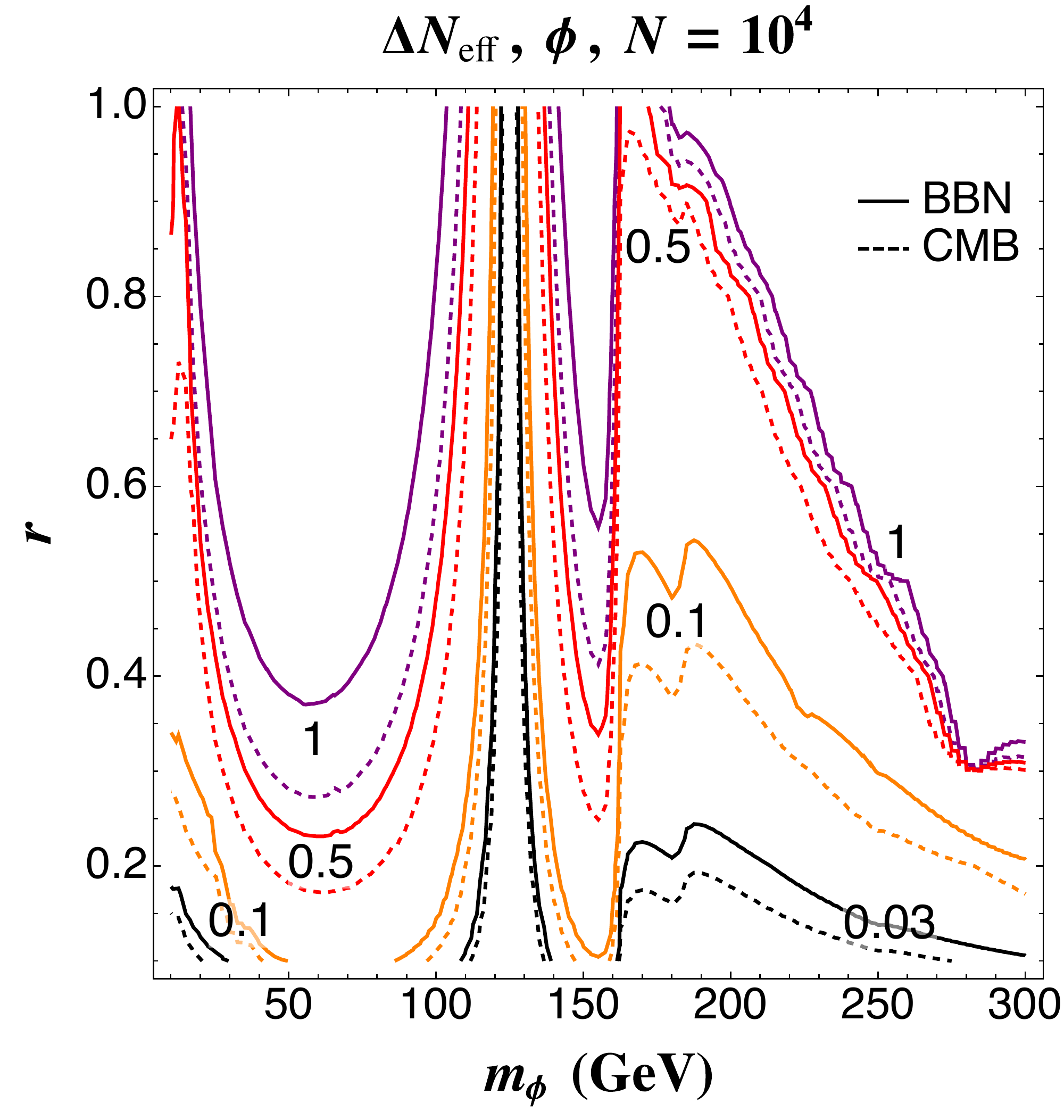} \quad\quad
\includegraphics[width=0.47\textwidth]{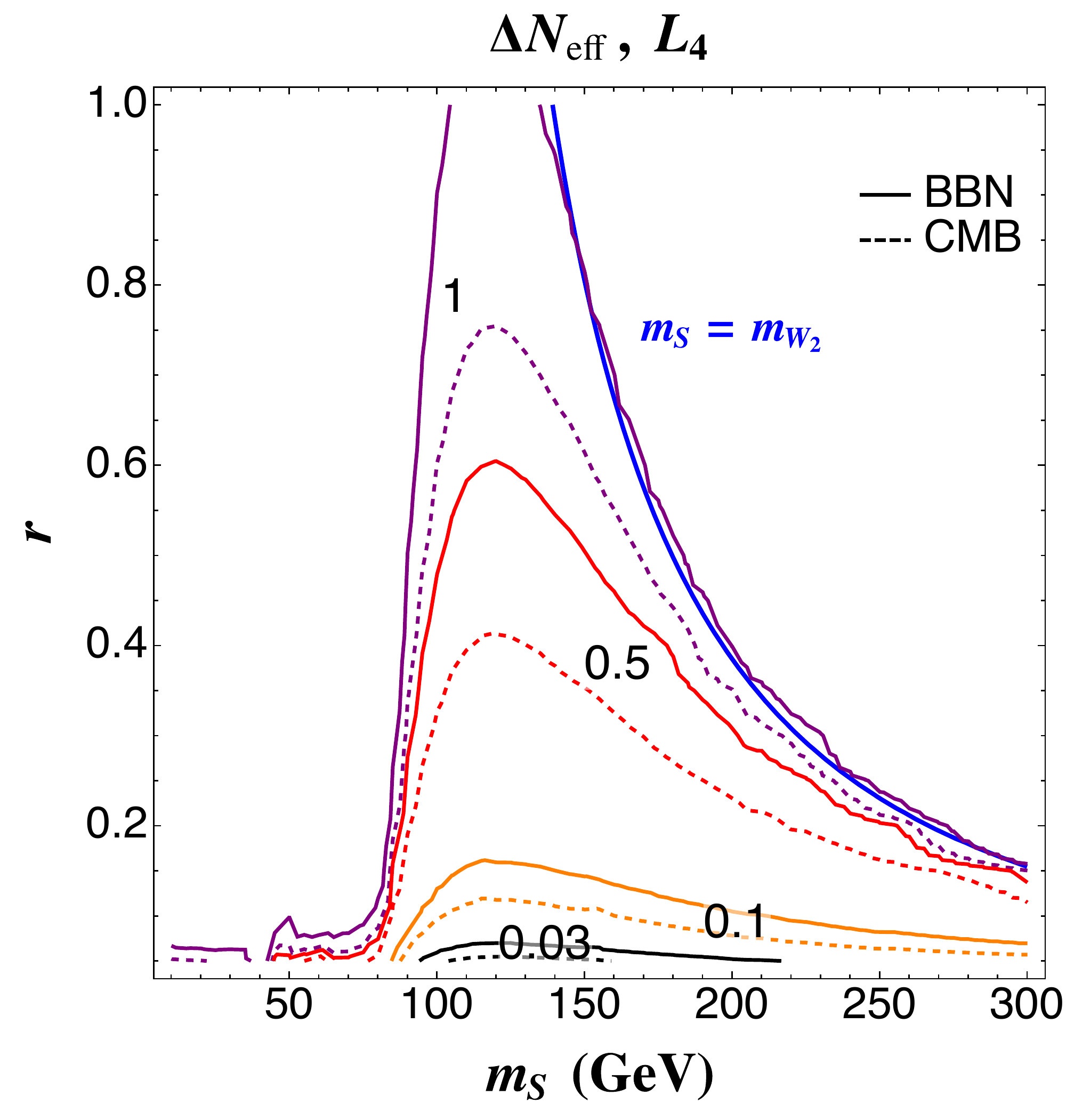}
\caption{$\Delta N_{\rm eff}$ contours as a function of reheaton mass and the $r$ parameter defined in Eq.~(\ref{eq:mHsqiScaling}). $\Delta N_{\rm eff}\simeq 0.03$ corresponds to the sensitivity of CMB stage 4 experiments. The current upper bound at the CMB epoch is around $0.6$. 
The left panel is for the $\phi$ model with $a=1$~MeV. The right panel is for the $L_4$ model with $\lambda\times \mu_E=1$~MeV, $M_L=400$~GeV, $M_{E, N}=500$~GeV, $Y_E=Y_N=0.2$, and $Y^c_E=Y^c_N=-0.5$.  As discussed in the text, the $L_4$ result is valid for a large range of $N$, namely $30 \lesssim N\lesssim 10^{9}$. Both figures were made using the zero temperature branching ratios of the reheaton; see the end of Sec.~\ref{Models} for a discussion.} 
\label{fig:Neff}
\end{center}
\end{figure*}
For example, assume that the reheaton is lighter than the lightest Higgs across all the sectors; then we have 
\begin{widetext}
\bea
\Delta N_{\rm eff}^{\phi} &\sim& {\displaystyle \sum_{i=1}^{N_b} \frac{1}{2\,i+1}} + \frac{y_c^2}{y_b^2}{\displaystyle \sum_{i=N_b+1}^{N_c} \frac{1}{2\,i+1}} \simeq \frac{1}{2}\left(\log 2N_b + \frac{y_c^2}{y_b^2}\log \frac{N_c}{N_b}\right)  \, , \quad\quad N_{b,c}=\left(\frac{m_{\phi}^2}{8\,m^2_{b,c}} -\frac{1}{2}\right)\, , \label{eq:Neff} \\
\Delta N_{\rm eff}^{L_4} &\sim& {\displaystyle \sum_{i=1}^{N_V} i^0} + {\displaystyle \sum_{i=N_V+1}^{N/2}}\frac{1}{(2\,i+1)^4} \simeq N_V\, , \quad\quad  N_V\simeq\left(\frac{M^2}{Y^2\, v^2} -\frac{1}{2}\right) \, , \nn
\eea
\end{widetext}
where $M$ represents one of the vector-like masses in the $L_4$ model and $Y$ one of the new Yukawas. In this estimate we have neglected the contribution from $m_H^2>0$ sectors and the effect of $g_*$ in each sector, to highlight the scaling of $\Delta N_{\rm eff}$. From this simple exercise we see that $\Delta N_{\rm eff}$ is dominated by the bottom of the spectrum. The sectors past $i=N_{b,c}$ or $i=N_V$ receive a negligible fraction of the total energy density and do not contribute to $\Delta N_{\rm eff}$. Using Eq.~(\ref{eq:Neff}) to go beyond a simple parametric estimate gives results that are in tension with current bounds. For example $m_\phi \simeq 50$~GeV implies $N_b \simeq 17$ and $\Delta N_{\rm eff} \simeq 2$. However, these estimates are only qualitative, and break down in a large fraction of the parameter space of the models. The results from a full numerical computation are shown in Fig.~\ref{fig:Neff}. 

There are two main messages that can be extracted from this calculation. First, we can satisfy current constraints for a range of reheaton masses up to a few hundred GeV. Second, the models predict values of $\Delta N_{\rm eff}$ within the range of sensitivity relevant for CMB stage 4 experiments~\cite{Dodelson:2013pln}. These next generation detectors, which should start taking data within the next five years, will probe $\Delta N_{\rm eff} \gtrsim 0.03$. If no beyond the SM discovery is made, then the only way to suppress this signal is to introduce ``fine tuning," which in the context of these models is the limit  $r \lesssim 0.1$. Alternatively, we could imagine alleviating this tension by taking the vector-like masses in the $L_4$ model far below the weak scale, in potential conflict with electroweak/Higgs measurements.  

A few additional features of the $\Delta N_{\rm eff}$ calculation are worth discussing.  In the $L_4$ case the plot is valid for a large range of $N$, namely $30 \lesssim N\lesssim 10^{9}$. The upper bound is determined by requiring $\lambda \lesssim 4\,\pi/\sqrt{N}$ and mixing between $e^c$ and the vector-like leptons less than 1\%. It is trivial to go beyond $N=10^{9}$, and even possible to reach $N = 10^{16}$, by lowering the reheaton coupling -- this comes at the expense of an overall decrease in reheating temperature, even though the result for $\Delta N_{\rm eff}$ would not change.  For $N < 30$, $\Delta N_{\rm eff}$ is smaller than shown in the figure. In the $\phi$ case, the results are more sensitive to $N$, as shown in Eq.~(\ref{eq:Neff}). We chose the largest $N$ that is both compatible with overclosure (see the next subsection) and also interesting from a model building perspective, given the relation to the Planck/GUT hierarchy ($N=10^4$). 

The shapes of the $\Delta N_{\rm eff}$ contours are easy to explain in terms of kinematics. In $L_4$ the allowed region corresponds to the reheaton decaying to our sector via a two-body channel, versus a three-body decay into all the other $m_H^2<0$ sectors. This is highlighted by the $m_S = m_{W_2}$ line in the plot. In the $\phi$ model the situation is different. The mixing with the Higgs naturally introduces a number of mass thresholds that reduce $\Delta N_{\rm eff}$.  At very low $\phi$ masses, decays to a pair of $b$-quarks are kinematically allowed only in our sector.  As the $\phi$ mass increases, the reheaton can mix resonantly with our Higgs and subsequently decay to a pair of $W$ or $Z$ bosons. The last aspect of these results that is not captured by the simple estimate in Eq.~(\ref{eq:Neff}) is the fact that $\left(\Delta N_{\rm eff}\right)_{\rm CMB} > \left(\Delta N_{\rm eff}\right)_{\rm BBN}$. It is easy to show that this must be the case by appealing to conservation of entropy in each of the sectors. If we compute the ratio of $\Delta N_{\rm eff}$ in sector $i$ at the two different epochs, we obtain
\begin{widetext}
\bea
\frac{\left(\Delta N_{\rm eff}^i\right)_{\rm CMB}}{\left(\Delta N_{\rm eff}^i\right)_{\rm BBN}} =\frac{g_{*}^i\left(T_{\rm CMB}^i\right)}{g_{*}^i\left(T_{\rm BBN}^i\right)} \left(\frac{g_{*S}^{\rm us}\left(T_{\rm BBN}^{\rm us}\right)}{g_{*S}^{\rm us}\left(T_{\rm CMB}^{\rm us}\right)}\right)^{4/3} 
 \left(\frac{g_{*S}^i\left(T_{\rm BBN}^i\right)}{g_{*S}^i\left(T_{\rm CMB}^i\right)}\frac{g_{*S}^{\rm us}\left(T_{\rm CMB}^{\rm us}\right)}{g_{*S}^{\rm us}\left(T_{\rm BBN}^{\rm us}\right)}\right)^{4/3} 
\simeq \left(\frac{g_{*S}^i\left(T_{\rm BBN}^i\right)}{g_{*S}^i\left(T_{\rm CMB}^i\right)}\right)^{1/3} \geq 1\, .
\eea
\end{widetext}
The first term in the first equality counts the number of relativistic degrees of freedom in sector $i$ at the two different temperatures. The second factor accounts for the fact that neutrinos in our sector are decoupled after BBN, so their temperature during the CMB epoch is lower than that of photons. The last term comes from entropy conservation in our sector and sector $i$. In the last equality we have used $g_{*}\simeq g_{*S}$.

To conclude the discussion of $\Delta N_{\rm eff}$, recall that the result depends almost exclusively on the reheaton branching ratios and that it is largely insensitive to the value of its overall coupling.  A single choice of $\lambda$ and $a$ is sufficient to understand the complete parameter space. In contrast, the precise value of the vector masses and Yukawa couplings in the $L_4$ model can change the results considerably, as it is already clear from Eq.~(\ref{eq:Neff}).  When the vector-like masses are around the TeV scale or above, the models are excluded, while $M\simeq 500$~GeV yields predictions that are consistent with current data, as shown in Fig.~\ref{fig:Neff}.  We leave a more detailed exploration of the parameter space and a discussion of possible collider signatures to future work. 

The second class of light particles that can impact our cosmological history are those that are non-relativistic at matter radiation equality, but might have free-streamed enough to suppress the matter power spectrum.
Particles that become non-relativistic at a time $t_\text{NR}<t_\text{EQ}$ suppress structure up to scales $\lambda_\text{FS} =c\, \sqrt{t_\text{EQ}^{\,} \,t_\text{NR}}(2+\log t_\text{EQ}/t_\text{NR})/a(t_\text{EQ})$. The neutrinos from many of the new sectors would have $\lambda_\text{FS}$ larger than one Mpc.  At these scales the matter power spectrum can be computed reliably in the linear regime and can be used to infer another upper bound on their energy density. To roughly estimate current constraints we compute the energy density in particles that can suppress structure at one Mpc or above. We find that for Dirac neutrinos the energy density is well below $1\%$ of the total dark matter energy density in all the plane of Fig.~\ref{fig:Neff} for both the $\phi$ and $L_4$ models, while for Majorana neutrinos this is true within the $\left(\Delta N_{\rm eff}\right)_\text{CMB}=0.5$ contours. 

The hot dark matter population may provide another signal.  The tower of sterile neutrinos results in a characteristic impact on the matter power spectrum. Furthermore, the hot dark matter signal is primarily determined by the reheaton branching ratios (and hence the spacing between the lightest sectors), so once a value of $\Delta N_{\rm eff}\neq 0$ is measured it is possible to make predictions for the distortion of the matter power spectrum and vice versa. In general our theories produce non-trivially related modifications in several CMB observables and we leave to future work a more detailed study. Our generic expectation is that neutrino cosmology is modified at the $\mathcal{O}(1)$ level due to slightly heavier albeit less abundant neutrinos in the closest sectors with electroweak symmetry breaking. 

\subsection{Massive stable particles}
Relic neutrinos account for a fraction $\Omega_\nu^\text{us} h^2 \simeq \sum m_\nu({\rm eV})/91.5 \gtrsim 10^{-3}$ of the energy density in the Universe. It is natural to ask if the heavier neutrinos in the sectors with $\vev{H}\neq 0$ can lead to overclosure problems. Furthermore, electrons and protons can be similarly problematic. This is perhaps surprising, since in the standard picture their symmetric component is completely negligible today. However, in the other sectors their masses are $\sqrt{i}$ larger and subsequently their annihilation cross-sections decrease as $1/i$.\footnote{For protons, this scaling is only valid once the quark masses exceed $\Lambda_{\text{QCD}}$; the scaling is slower for the nearer sectors.} 

In all cases, the relic density of the new stable particles comes from two different sources. There is a contribution that grows with $i$ from the sectors where the stable particles reach thermal equilibrium (including a possible freeze-in abundance from our sector) and a second contribution that decreases with $i$ from sectors where the particles never thermalize.  Let us focus on this first contribution (for the moment we will neglect the freeze-in abundance from our sector):
\bea
\Omega\, h^2 &=& \frac{s_0}{\rho_c^0}\sum_{i=-1}^{-N_d} m_i\, Y_i^{\text{fo}} +... = a \left(N_d\right)^p+...\, .
\eea
Here we use $\Omega\, h^2$ to indicate the relic density of either neutrinos, electrons or protons; $\rho_c^0$ is the critical energy density today; $s_0$ is the entropy density; $m_i$ is the mass of the stable particle; $Y_i^{\text{fo}}$ is its yield at freeze-out; $N_d$ is the sector after which the stable particles are not ever in thermal equilibrium with the other particles in their sector; and $a$ and $p$ are positive numbers. In general $a \sim \Omega^{\text{us}} h^2$ and $p>1$. The reason for $p>1$ is that $m_i\sim \sqrt{i}$ (or $\sim i$ for Majorana neutrinos) and up to a certain sector number $Y_i^{\text{fo}}$ also grows with $i$, since neutrinos, electrons and protons all freeze-out earlier and earlier. 

In the $\phi$ model this thermal abundance turns out to be the only relevant one.  Specifically, electrons and positrons provide the dominant constraint. Once the bound on the reheating temperature is taken into account, the freeze-in abundance from our sector is negligible. Furthermore the overclosure bound on $N$ kicks in before including heavy enough sectors where electrons would not thermalize. Therefore the bound arises only from thermal freeze-out ($n_{e}^i \sim 1/\langle \sigma_e v\rangle_i$) and it is straightforward to estimate:
\bea
\Omega_e^\phi\, h^2 &=& \sum_{i=1}^{N_\text{th}} \frac{m_{e}^i\, n_{e}^i}{\rho_c^0} \simeq \frac{\big(m_e^{\rm us}\, T_0^{\rm us}\big)^3}{\rho_c^0}\frac{N^{5/2}_\text{th}}{M_\text{pl}\, v_{\text{us}} \,\alpha^2}\,  \notag\\
 &\lesssim& 0.1\times \Omega_\text{DM}\, h^2 \quad \Longrightarrow \quad N_\phi\lesssim 10^5 \, ,
\eea
where the sum runs up to the heaviest sector where the electrons have thermalized as denoted by $N_\text{th}$, $T_0^{\rm us}$ is the photon temperature in our sector today, $m_e^{\rm us}$ is our electron mass, $M_\text{pl}$ is the Planck mass, and all other quantities were defined previously.  For this estimate we have assumed that our sector dominates the energy density of the Universe when electron-positron annihilations freeze-out, \emph{i.e.}, the $\Delta N_{\rm eff}$ constraint is satisfied. Furthermore we have conservatively assumed freeze-out happens just after reheating (at $T_\text{RH}^{\rm us}\simeq v_{\text{us}}$) in all the sectors. 
Finally, note that we have required that electrons and positrons make up only $10\%$ of dark matter, the rough bound for particles that behave very differently from cold, collisionless dark matter. To be more conservative we could require them to make up only $1\%$ of dark matter, which would reduce the maximum allowed value of $N_\phi$ by $60\%$, still leaving open $N_\phi=10^4$.
 
To conclude this section we note that the rapid scaling of the energy density with sector number in $L_4$ protects the model from overclosure.  This implies that $N$ can be taken all the way to $10^{16}$ and still be consistent with data, at the expense of a low reheat temperature.

\subsection{Mixing Between Sectors}
\label{sec:MixingBetweenSectors}
Upon integrating out the reheaton, the low energy theory will contain cross-couplings between the sectors.  Stringent bounds from stellar and supernova cooling place limits on the size of these mixings.

\paragraph{\boldmath $\epsilon_i \,F_{\mu \nu}\, F_i^{\mu \nu}$}

In the presence of kinetic mixing, the electrically charged particles of other sectors will have milli-charge couplings to our photon.  The most stringent bound on this coupling is derived from energy loss in stars~\cite{Davidson:1991si, Davidson:2000hf}.  In sectors with $m_H^2 > 0$, the charged particles are all extremely light -- much lighter than stellar temperatures -- so that democratic kinetic mixing leads to $\mathcal{O}(N)$ accessible final states for plasmon decay.  Thus we require $\sqrt{\sum_i \epsilon_i^2} \lesssim 10^{-14}$, in which $\epsilon_i$ is the coefficient of kinetic mixing between our photon and that of sector $i$, and $i$ runs over all sectors with $m_H^2 > 0$.

Accordingly, there must be no bi-fundamental matter in the UV, the inclusion of which would generate kinetic mixing at one loop.  Even in the absence of those states, kinetic mixing may be generated in the IR through the coupling to the reheaton.\footnote{Kinetic mixing is not generated at any order if the coupling between sectors is mediated by a single, real scalar field, since there can be no effective coupling of such a field to the electromagnetic field-strength tensor.  Therefore, this effect can be safely neglected in the $\phi$ model.}  In this case the bounds may be easily avoided by the smallness of the coupling.   As described in Sec.~II, the portal couplings must decrease with increasing $N$ in order to have a consistent large-$N$ limit.  

For example, in the $L_4$ model, we must have $\lambda \sim \lambda_0 / \sqrt{N}$.  The kinetic mixing parameter is generated only at three loops with four powers of the portal coupling:
\begin{align}
\epsilon_i \sim \frac{\alpha}{4\, \pi}\left(\frac{\lambda_0}{4\,\pi}\right)^4 \frac{1}{N^2 \,i}.
\end{align}
Note that $\epsilon_i$ decreases with $i$ due to the scaling of $\big(m_H^2\big)_i$, and so kinetic mixing is dominated by the sectors nearest to our own.  Then the stellar bounds may be avoided as long as $\lambda_0 / 4\pi \lesssim 10^{-3} \sqrt{N}$, and no suppression is required for $N \gtrsim 10^6$ beyond the natural large-$N$ scaling.

\paragraph{\boldmath $\epsilon_i^n\, \nu_i^{\dagger}\, \bar{\sigma}^{\mu} D_{\mu}\, \nu$}
At one loop in the $L_4$ model, the reheaton mass-mixes with neutrinos.  After integrating out the reheaton, this induces kinetic mixing between neutrinos of different sectors.  However, because the vector-like leptons only couple to the charged leptons, the effective coupling to the neutrinos is Yukawa-suppressed:
\begin{align}
\epsilon^n_i \sim (m_{\ell})_{\text{us}} \,(m_{\ell})_i \left(\frac{\lambda\, Y_E^c\, \mu_E}{16\, \pi^2 \,(M_4)_i \,m_S} \right)^2.
\end{align} 
For sectors with $(M_4)_i \sim M_L$, $\epsilon_i \sim \sqrt{i}$.  Once $Y_E \,v_i \gtrsim M_L$, the kinetic mixing decreases as $\epsilon_i \sim 1/\sqrt{i}$.  

Energy loss in SN1987a~\cite{Raffelt:1987yt} limits the size of the kinetic mixing.  The neutrino production rate from neutral-current bremsstrahlung requires $\sqrt{\sum_i (\epsilon^n_i)^2} \lesssim 10^{-4}$.  Due to the growth of $\epsilon^n_i$ with $i$ for small values of $i$, the sum is dominated by those sectors for which the vector-like lepton masses are larger than their chiral masses.  For typical parameters, such as those shown in Fig.~\ref{fig:Neff}, this is the case for only $\mathcal{O}(10)$ sectors.  Taking into account the bound on $T_{\text{RH}}$ from Sec.~\ref{Models}, this gives $\sqrt{\sum_i (\epsilon^n_i)^2} \lesssim 10^{-13} \left(M_L / 4\, \pi\, v \right)^4$, so that there is no constraint as long as the vector-like masses are taken to be sufficiently close to the weak scale.

\paragraph{\boldmath $\epsilon^c_i \,G_F\, (\nu_i^{\dagger}\, \bar{\sigma}^{\mu} \,e^c)\left(p^{\dagger}\, \bar{\sigma}_{\mu}\, n\right)$}
There is a somewhat more powerful constraint from SN1987a due to charge-current neutrino production.  The mass-mixing of the reheaton with neutrinos leads to an effective four-Fermi operator, with
\begin{align}
\epsilon^c_i \sim (m_{\tau} \,m_{\nu})_i \left(\frac{\lambda\, Y_E^c \,v_\text{us}\, \mu_E\, M_L}{4\,\pi\, m_S \,(M_4^2)_{\text{us}} \,(M_4)_i}\right)^2.
\end{align}
In the case of Majorana neutrino masses, $\epsilon_i^c$ grows like $\sqrt{i}$, so that $\sqrt{\sum_i (\epsilon_i^{c})^2} \sim N$.  Once again taking into account the limit on $T_{\text{RH}}$, we have  $\sqrt{\sum_i (\epsilon_i^{c})^2}~\lesssim~10^{-24}\ N\, \big(M_L / \sqrt{4\, \pi\, Y_E} \,v\big)^4$.  The supernova bound is only $\sim 10^{-5}$, so that even for $N \sim 10^{16}$, the coupling is unconstrained for $M_L$ near the weak scale.

Finally, limits on active-sterile neutrino oscillations can also bound $\epsilon^c_i$, both from cosmological measurements as well as active neutrino disappearance~\cite{Cirelli:2004cz}.  However, due to the Yukawa suppression of the neutrino mixing, the most relevant limits are those involving the tau neutrino, which are comparatively weak.  Absent resonant mixing due to accidental degeneracies, which we expect to be atypical in our parameter space, the bounds from neutrino oscillations are negligible.

\subsection{Colliders}
\label{sec:Colliders}
Models of $N$naturalness can provide collider signatures through both direct production of the reheaton as well as rare decays of SM particles.  However, the smallness of the reheaton couplings, due to both large-$N$ suppression and $T_{\text{RH}}$ constraints, precludes these signatures from being a generic feature of our models.  In the $\phi$ model, for example, rare Higgs decays proceed through $\phi - h$ mixing.  The dominant signature in this case is are invisible decays of the SM-like Higgs boson, with $\text{BR}_\text{inv} \sim \theta_{\phi h}^2 \,\Delta N_{\text{eff}} / (1 + \Delta N_{\text{eff}}) \lesssim 10^{-12} $ after requiring sufficiently low $T_{\text{RH}}$.  Even using optimistic estimates, future colliders such as TLEP or a 100~TeV machine (with 10 ab$^{-1}$ of luminosity) will only produce $\sim 10^6$~\cite{Gomez-Ceballos:2013zzn} or $10^{10}$~\cite{Arkani-Hamed:2015vfh, Golling:2016gvc} Higgses, respectively, rendering such decays unobservable.

Direct production of $\phi$ is similarly suppressed, since it must proceed through the same mixing angle.  Even for $m_{\phi} < m_h$, in which case the production cross sections are somewhat larger, direct production of $\phi$ will be unobservable.  For example, a SM-like Higgs with a mass of 10~GeV gives a cross section only 2.5 times larger at TLEP and approximately 14 times larger at a 100~TeV $p-p$ machine than a Higgs at 125~GeV.  Nevertheless, if $\phi$ is sufficiently light, there may be a variety of interesting signatures; see, {\emph e.g.},~\cite{Clarke:2013aya} for a study of current constraints which probe mixing angles down to $\theta_{\phi h} \sim 10^{-5}$.  We leave a detailed study of detection prospects for $m_{\phi} \lesssim 10$~GeV to future work, see \emph{e.g.}, \cite{Krnjaic:2015mbs} for an analysis of a similar scenario.

In the $L_4$ model, the new sectors and the reheaton are similarly difficult to observe; however, the vector-like leptons of our own sector may be accessible.  As discussed above, the vector-like mass parameters should all be of order the weak scale.  This implies they would likely be observable, both directly and through $h\to \gamma\, \gamma$ and precision electroweak measurements.  For a recent study of these bounds, see~\cite{Altmannshofer:2013zba}.  In particular, attempting to evade these constraints by raising the mass of the vector-like leptons will re-introduce tension with $\Delta N_{\text{eff}}$, as described in the Section~\ref{subsec:massless}.

\subsection{A Heavy Axion}
\label{sec:HAxion}

An outstanding puzzle within the SM is the strong CP problem.  
If we have $N$ copies of the SM, then naively they all have their own theta angles $\theta_i$.  However, if the $S_N$ symmetry is only softly broken by Higgs mass terms, then all of these angles would be equal.  A shared axion would be able to set to zero all the $\theta_i$'s at the same time.  The only difference between the version proposed here and the single sector story is that here there are $N$ contributions to its mass from each $\Lambda_\text{QCD}$.  The potential for the axion has three contributions
\begin{align}
V(a) \ni 
\left\{
\begin{array}{ll}
\frac{\Lambda_{\text{QCD},i}^6}{m_{H,i}^2} \left(\frac{a}{f_a} - \theta_i\right)^2  &  \text{for } i < 0   \\[8pt] 
 m_{\pi,i}^2\, f_{\pi,i}^2 \left(\frac{a}{f_a} - \theta_i\right)^2 & \text{for } 0 \leq i < N_u\\[8pt]
  \Lambda_{\text{QCD},i}^4 \left(\frac{a}{f_a} - \theta_i\right)^2 & \text{for }  i \geq N_u
\end{array}
\right.\,,
\label{Eq: axion}
\end{align}
where $\Lambda_{\text{QCD},i}$ is the QCD scale for the $i^\text{th}$ sector, $i<0$ corresponds to sectors with $\vev{H_i} = 0$, and $N_u \sim 10^5$ is the sector with the smallest vev for which $m_u > \Lambda_\text{QCD}$.  The contribution from the sectors with $\vev{H_i} = 0$ are due to higher dimensional operators from integrating out the Higgs doublet, the sectors with $m_u < \Lambda_{\text{QCD}}$ yield the familiar contribution to the axion potential, and the final term is the result for pure QCD with no light quarks.  Numerically, the first term can always be neglected, the second term dominates as long as $N < N_u$, and only the third term is relevant for $N \gg N_u$.

In order to estimate how much heavier this state will be as compared to the standard case, we have calculated Eq.~(\ref{Eq: axion}) numerically including the one-loop running of $\Lambda_\text{QCD}$.  For the first two sums in Eq.~(\ref{Eq: axion}), we used chiral perturbation theory to calculate the contribution to the axion mass.  For the last sum, we normalized it such that it is equal to the chiral perturbation theory result when $m_u = \Lambda_\text{QCD}$.  A numerical fit to the axion mass  gives approximately
\begin{align}
\frac{m_a(N)}{m_a(1)} \simeq \left\{
\begin{array}{ll}
4 \times 10^3 \left(\frac{N}{10^4}\right)^{1} \quad\quad\quad& \text{for } N < N_u \\[2pt]
 2 \times 10^{14} \left(\frac{N}{10^{16}}\right)^{0.9}\, & \text{for } N \gg N_u \\
\end{array}
\right.\, .
\end{align}

It is critical that the soft-breaking of the $S_N$ symmetry by the different Higgs vevs does not lead to any issues via higher dimensional operators.  For example, one class of operators that leads to a change in $\theta_i$ between the sectors are 
\bea
\mathcal{O}_{\Delta\theta_\text{QCD}} \sim Y_u\, H_i\, Q_i\, u_i^c \frac{|H^i|^2}{\Lambda_G^2}\,.
\eea
Because the different sectors have different Higgs vevs, a chiral rotation shows that the theta angles all differ by $\sim |H_i|^2/\Lambda_G^2$.  Plugging this into Eq.~(\ref{Eq: axion}), solving for the axion vev, and requiring that our theta angle is smaller than $10^{-10}$, we find that $N < 10^{10}$ if a shared axion is the solution to the strong CP problem.  This approach requires the important assumption that whatever resolves the hierarchy problem between $\Lambda_H$ and $\Lambda_G$ does not introduce these operators or any other Higgs dependent phases.


\section{Discussion}
\label{sec:Discussion}

In this paper we have proposed a new solution to the hierarchy problem. The need for a huge integer $N$ is obviously the least appealing feature of our setup. It is perhaps not entirely unreasonable to have the mild $N \sim 10^4$ compatible with the existence of a supersymmetric GUT scale, but this seems outlandish in the $N \sim 10^{16}$ limit. At the moment it is difficult to see how such a large integer can be explained dynamically, in the same way as we usually explain hierarchies by, \emph{e.g.} dimensional transmutation. On the other hand, this is simply another large set of degrees of freedom, and we do not deeply understand where the even vaster number of degrees of freedom in a macroscopic expanding universe comes from, so perhaps the large $N$ may eventually find a different sort of natural explanation. The theoretical consistency of the proposal also makes a number of demands on the UV theory, such as the absence of sizable cross-couplings between the sectors, which may be technically natural but may again strain credulity.  However, we find it fascinating that huge values of $N$ are experimentally viable. This is highly non-trivial, and indeed in the simplest models we did find significant constraints on $N$. While we have examined all the zeroth-order phenomenological constraints we know of, it is important to continue to look for constraints on (and signals of!)\! the scenarios with high values of $N \big(\gg 10^4\big)$.

It is also interesting to compare $N$naturalness with other approaches. It bears a superficial resemblance to large extra dimensions, which add $10^{32}$ degrees of freedom in the form of KK gravitons, as well as the scenario of Dvali~\cite{Dvali:2007hz} which invokes $10^{32}$ copies of the SM.  In each of these cases, $M_{\text{pl}}$ is renormalized down to the TeV~scale.  Of course this predicts (as yet unseen) new particles accessible to the LHC~\cite{Dvali:2009ne}.  By contrast, $N$naturalness solves the hierarchy problem with cosmological dynamics; the weak scale is parametrically removed from the cutoff, and so it does not demand new physics to be accessible at colliders.

$N$naturalness has some features in common with low-energy SUSY as well. Both models invoke a softly broken symmetry: SUSY is broken by soft terms, and the $S_N$ symmetry is broken by varying Higgs masses.  Also in both cases, the most obvious implementations of the idea are experimentally excluded.  If SUSY is directly broken in the MSSM sector, we have the famous difficulties with charge and color breaking; in the case of $N$naturalness, direct reheating of all $N$ sectors is grossly excluded by $N_\text{eff}$.  Thus in both cases we need to have ``mediators.''  SUSY must be dominantly broken in another sector and have its effects mediated to the MSSM.  Similarly, reheating must be dominantly communicated to the reheaton, which subsequently dumps its energy density into the other sectors.  Finally, both models have additional scales that are not, on the face of it, tied to the physics responsible for naturalness.  In SUSY there is a ``$\mu$ problem'' in that the vector-like Higgsino mass must be comparable to the soft scalar masses, while in $N$naturalness the reheaton mass must be close to the bottom of the spectrum of Higgs masses.  While in both cases there are simple pictures for how this can come about, these coincidences do not emerge automatically.

Moving beyond purely field theoretic mechanisms, there is the recent proposal of the relaxion~\cite{Graham:2015cka}, which invokes an extremely long period of inflation coupled with axionic dynamics to relax to a low weak scale. While both the relaxion and $N$naturalness mechanisms are cosmological, the physical mechanism of the relaxation, associated with the huge number of $e$-foldings of inflation, is {\it in principle} unobservable given our current accelerating Universe, much like the vast regions of the multiverse outside our cosmological horizon are imperceptible. By contrast, the cosmological dynamics associated with reheaton decay in $N$naturalness are sharply imprinted on the particle number abundance in all the sectors.  They are not only in principle observable but, as we have stressed (at least for a small number of sectors ``close'' to ours), are detectable in practice within our Universe.

It is also interesting to contrast $N$naturalness with the picture of an eternally inflating multiverse, with environmental selection explaining the smallness of the cosmological constant, as well as potentially at least part of the hierarchy problem. This picture is, after all, the first cosmological approach to fine-tuning puzzles.  While it is very far from well-understood and has yet to make internal theoretical sense, it is the only cartoon we have for understanding the cosmological constant problem and does not involve any model-building gymnastics.  Furthermore, fine-tuning for the Higgs mass also has a plausible environmental explanation. Especially in the context of minimal split SUSY~\cite{ArkaniHamed:2004fb}, these ideas give us a picture which simultaneously accounts for the apparent fine-tuning of the cosmological constant and the Higgs mass, while maintaining the striking quantitative successes of natural SUSY theories in the form of gauge coupling unification and dark matter.  Nonetheless, it is important to continue to look for alternatives, minimally as a foil to the landscape paradigm.  $N$naturalness is a concrete example of an entirely different cosmological approach to tuning puzzles, and in particular relies on the existence of only a single vacuum.  

We note that there is no obstacle to augmenting $N$naturalness with an anthropic solution to the cosmological constant problem. The presence of extra sectors exponentially increases the number of available vacua. For example we could add to the SM a sector with $m$ vacua and end up with $m^N$. Already $N\simeq 10^4$ with two vacua per sector is more than enough to scan the cosmological constant without relying on string theory landscapes.  When solving the entire hierarchy problem with $N \simeq 10^{16}$, the vacua utilized to scan the cosmological constant can even be the two minima of the Higgs potential; this requires a high cutoff so that the second minimum is below $\Lambda_H$ and the difference in the potential energy of the two minima is $\mathcal{O}\big(\Lambda_G\big)$.

To conclude, we would like to comment on the nature of the signals that we have discussed in this paper.  For concreteness, three models that make $N$naturalness cosmologically viable were presented.  However, it is easy to imagine a broader class of theories that realizes the same mechanism. We can relax the assumption that the Higgs masses are uniformly spaced (or even pulled from a uniformly distribution) or that all the new sectors are exact copies of the SM. It is also possible to construct different models of reheating, with new physics near the weak scale to modify the UV behavior of the theory.   

Nonetheless our sector can not be special in any way.  There will always be a large number of other sectors with massless particles and with matter and gauge contents similar to ours, leading to the following signatures: 
\mbox{} \\
\begin{itemize}[leftmargin=12pt]
\item We expect extra radiation to be observable at future CMB experiments. 
\item The neutrinos in the closest $m_H^2<0$ sectors are slightly heavier and slightly less abundant than ours. This implies $\mathcal{O}(1)$ changes in neutrino cosmology, which will start to be probed at this level in the next generation of CMB experiments~\cite{Allison:2015qca}. 
\item If the strong CP problem is solved by an axion, its mass will be much larger than the standard prediction.  
\item If $N\lesssim 10^4$ as motivated by grand unification, supersymmetry or new natural dynamics should appear beneath $10$~TeV.
\end{itemize}
\mbox{} \\

The natural parameter space is being probed now, and soon we may know if the $N$naturalness paradigm explains how the hierarchy problem has been solved by nature.

\vspace{.3in}
\section*{Acknowledgments}
\vspace{-0.1in}
We are grateful to Bob Holdom, Jared Kaplan, Marilena LoVerde and Kris Sigurdson for useful discussions. We also want to thank Arka Banerjee and Neal Dalal for a preliminary investigation of the neutrino cosmology. NAH is supported in part by U.S. Department of Energy grant de-sc0009988. TC is supported by an LHC Theory Initiative Postdoctoral Fellowship, under the National Science Foundation grant PHY-0969510.  RTD acknowledges support from the Institute for Advanced Study Marvin L. Goldberger Membership and DOE grant de-sc0009988.  AH is supported by the Department of Energy grant DE-SC0012012 and National Science Foundation grant 1316699.  HDK is supported by National Research Foundation of Korea(NRF) No. 0426-20140009 and No. 0409-20150110.  TC and AH acknowledge the Aspen Center for Physics where this work was completed, which is supported by National Science Foundation grant PHY-1066293.

\end{spacing}

\bibliography{ref}
\bibliographystyle{utphys}

\end{document}